\documentstyle[aps,preprint]{revtex}
\begin{document}
\draft
%

\title{Superheavy nuclei in relativistic effective Lagrangian model} 

\author{Tapas Sil$^1$, S. K. Patra$^2$, B. K. Sharma$^2$,
        M. Centelles$^1$, and X. Vi\~nas$^1$}

\address{
$^1${\it Departament d'Estructura i Constituents de la Mat\`eria,
     Facultat de F\'{\i}sica,
\\   Universitat de Barcelona,
     Diagonal {\sl 647}, {\sl 08028} Barcelona, Spain}
\\
$^2${\it Institute of Physics, Sachivalaya Marg, 
         Bhubaneswar {\sl 751 005}, India.}
}


\maketitle

\begin{abstract}
Isotopic and isotonic chains of superheavy nuclei are analyzed to
search for spherical double shell closures beyond $Z=82$ and $N=126$
within the new effective field theory model of Furnstahl, Serot, and
Tang for the relativistic nuclear many-body problem. We take into
account several indicators to identify the occurrence of possible
shell closures, such as two-nucleon separation energies, two-nucleon
shell gaps, average pairing gaps, and the shell correction energy. The
effective Lagrangian model predicts $N=172$ and $Z=120$ and $N=258$
and $Z=120$ as spherical doubly magic superheavy nuclei, whereas
$N=184$ and $Z=114$ show some magic character depending on the
parameter set. The magicity of a particular neutron (proton) number in
the analyzed mass region is found to depend on the number of protons
(neutrons) present in the nucleus.

\end{abstract}

\vspace*{0.5cm}


\vspace*{0.5cm}


\newpage

\section{Introduction}

In the last thirty years a continuing effort has been devoted to the
investigation of superheavy nuclei both in experiments and in
theoretical research. A fascinating challenge in the study of these
nuclei is the quest for the islands of stability where the next magic
numbers beyond $N=126$ and $Z=82$ may be located. Experiments made at
GSI, Dubna and Berkeley have allowed the synthesis and detection of
some superheavy nuclei. For instance, light isotopes of the elements
$Z=110$, 111 and 112 have been obtained at GSI and Dubna
\cite{hof,hof2,arm}. They have been identified by their characteristic
$\alpha$-decay chains which lead to already known isotopes. These new
nuclei are expected to be deformed, consistently with the predicted
occurrence of a deformed magic shell closure at $Z=108$ and $N=162$
(see e.g.\ Refs.\ \cite{mol,smo,bur}). First data of some heavier and
more neutron-rich isotopes of atomic number $Z=112$ ($N=171$), $Z=114$
($N=173$--175) and $Z=116$ ($N=176$) produced by means of fusion
reactions have also been measured at Dubna \cite{oga}.

Theoretical predictions made at the end of the sixties pointed towards
the existence of an island of long-lived superheavy elements (SHE)
centered around $N=184$ and $Z=114$ \cite{mye,sob,nil,mos}. The nuclei
around the hypothetical doubly magic element $^{298}$114 were expected
to be nearly spherical with longer half-lives. Such superheavy nuclei,
having a negligible liquid-drop fission barrier, would be stabilised
mostly by quantal shell effects. Many of the more recent theoretical
works on superheavy nuclei are based on the nuclear mean field
approach and can be classified in two main groups. On the one hand, we
have the macroscopic-microscopic models which include a liquid-drop
contribution for the part of the energy which varies smoothly with the
number $A$ of nucleons, and a shell correction contribution obtained
from a suitable single-particle potential for the fine tuning. On the
other hand, there are the self-consistent Hartree-Fock or Hartree
calculations based on Skyrme forces or on the relativistic non-linear
$\sigma-\omega$ model, respectively.

The nuclei in the range around $Z\approx 110$ already detected in
experiments bridge the gap between the known actinides and the unknown
superheavy elements. With the advent of more experimental data, a
commendable endeavor has been undertaken in nuclear structure research
\cite{bur,cwiok99,bender00,pat,mamdouh01,ren_npa1,ren_prc1,ren_prc2,ren_prc3,nazar02,cranked03,typel03,bender03,gambhir03,bur04} 
aimed at verifying the reliability of the present theoretical models
in the regime of the heavier actinides and of the discovered
superheavy nuclei around $Z=110$, which requires deformed
calculations. The fact that many of the observed data for SHE are for
odd-even decay chains renders the calculations and the comparison with
experiment even more complicated, since the deformed level density is
high and the observed nuclei may be in isomeric states. Calculations
with self-consistent models of some $\alpha$-decay chains
\cite{nazar02}, deformation energy curves along the fission path
\cite{bur04}, and shell structures \cite{bur} find that there is a
gradual transition from well-deformed nuclei around the deformed
$Z=108$ and $N=162$ shell closures to spherical shapes approaching
larger superheavy nuclei around the putative $N=184$ magic neutron
gap, in qualitative agreement with the earlier studies in mac-mic and
semiclassical models. Still, the Hartree-Fock model mass formula of
Ref.\ \cite{goriely02} predicts large deformations in many of the
isotopes of $Z=114$ and in almost all of the $Z=120$ isotopes. As
pointed out in some recent works, the description of deformed SHE may
require to consider triaxial deformations and reflection-asymmetric
shapes \cite{bur04,cwiok96,bender98} (Ref.\ \cite{hirata96} pioneered
the relativistic mean field triaxial calculations). It is even
possible that there exist isolated islands of stability associated
with exotic (semi-bubble, bubble, toroidal, and band-like) topologies
in nuclei with very large atomic numbers
\cite{nazar02,ber,decharge03}. 

Another longstanding goal of the nuclear structure studies in the
field of superheavy nuclei has been to establish the location in 
$N$ and $Z$ of the next {\em spherical double} shell closures
for elements heavier than $^{208}$Pb, and of the largest shell effects
which are a necessary condition for the stability of SHE against
fission. In this context, most of the calculations published in the
literature are performed in spherical symmetry. It is well established
that the macroscopic-microscopic calculations predict spherical shell
closures at $Z=114$ and $N=184$ \cite{mol}. In self-consistent
calculations, however, the proton and neutron shell structures
strongly affect each other and other $N$ and $Z$ values can appear as
candidates for shell closures depending on the model interaction. For
example, Hartree-Fock calculations with a variety of Skyrme forces
show the most pronounced spherical shell effects at $Z=124,126$ and
$N=184$ \cite{rut,ben,kru,rei,ben2}. As an exception to this rule,
Skyrme parametrizations such as SkI3 and SkI4 which have a modified
spin-orbit interaction prefer $Z=120$ and $Z=114$, respectively, for
the proton shell closure \cite{ben,kru}. Hartree-Fock-Bogoliubov
calculations with the finite range Gogny force predict $Z=120,126$ and
$N=172,184$ as possible spherical (or nearly spherical) shell closures
\cite{ber,decharge03}. At variance with Skyrme Hartre--Fock, the
relativistic mean field (RMF) theory with the conventional scalar and
vector meson field couplings typically prefers $Z=120$ and $N=172$ as
the best candidates for spherical shell closures
\cite{rut,ben,kru,rei,ben2}. Of course, the different nature of the
spin-orbit interaction in the Skyrme and RMF models is pivotal in
deciding the location of the stronger shell effects. Detailed
comparisons between Skyrme Hartree-Fock and RMF calculations of SHE
can be found in Refs.\ \cite{ben} and \cite{rei}.

The discrepancies in the predicted spherical shell closures for SHE
motivate us to reinvestigate them using the more general RMF model
derived from the chiral effective Lagrangian proposed by Furnstahl,
Serot and Tang \cite{fur,fur2,ser2}. In this first attempt to apply
the effective field theory (EFT) model to the region of superheavy
nuclei we will restrict ourselves to analyze the occurrence of double
shell closures and the shell stabilizing effect in spherical symmetry,
as done e.g.\ in Refs.\ \cite{rut,ben,kru,rei,ben2}. We want to learn
whether in this respect the EFT approach shows a different nature
compared to the usual RMF theory or not. The possible extension of the
calculations to deformed geometries and the subsequent application of
the EFT model to the heavier actinides and the lighter transactinides
where experimental data have been measured is left for future
consideration.

The relativistic model of Refs.\ \cite{fur,fur2,ser2} is a new
approach to the nuclear many-body problem which combines the modern
concepts of effective field theory and density functional theory (DFT)
for hadrons. An EFT assumes that there exist natural scales to a given
problem and that the only degrees of freedom relevant for its
description are those which can unravel the dynamics at the scale
concerned. The unresolved dynamics corresponding to heavier degrees of
freedom is encoded in the coupling constants of the theory, which are
determined by fitting them to known experimental data. The Lagrangian
of Furnstahl, Serot and Tang is intended as an EFT of low-energy
QCD\@. As such, its main ingredients are the lowest-lying hadronic
degrees of freedom and it has to incorporate all the infinite (in
general non-renormalizable) couplings consistent with the underlying
symmetries of QCD\@. To endow the model with predictive power the
Lagrangian is expanded and truncated. Terms that contribute at the
same level are grouped together with the guidance of naive dimensional
analysis. Truncation at a certain order of accuracy is
consistent only if the coupling constants eventually exhibit
naturalness (i.e., if they are of order unity when in appropriate
dimensionless form). In the nuclear structure problem the basic
expansion parameters are the ratios of the scalar and vector meson
fields and of the Fermi momentum to the nucleon mass $M$, as these
ratios are small in normal situations. To solve the equations of
motion that stem from the constructed effective Lagrangian one applies
the relativistic mean field approximation in which the meson fields
are replaced by their classical expectation values.

EFT and DFT are bridged by interpreting the expansion of the effective
Lagrangian as equivalent to an expansion of the energy functional of
the many-nucleon system in terms of nucleon densities and auxiliary
meson fields. The RMF theory is then viewed as a covariant formulation
of DFT in the sense of Kohn and Sham \cite{spe}. That is, the mean
field model approximates the exact, unknown energy functional of the
ground-state densities of the nucleonic system, which includes all
higher-order correlations, using powers of auxiliary classical meson
fields. This merger of EFT and DFT provides an approach to the nuclear
problem which retains the simplicity of solving variational Hartree
equations with the bonus that further contributions, at the mean field
level or beyond, can be incorporated in a systematic and controlled
manner. 

If the chiral effective Lagrangian is truncated at fourth order, in
mean field approach one recovers the same couplings of the usual
non-linear $\sigma-\omega$ model plus additional non-linear
scalar-vector and vector-vector meson interactions, besides tensor
couplings \cite{fur2,ser2}. The free parameters of the resulting
energy functional have been fitted to ground-state observables of a
few doubly-magic nuclei. The fits, parameter sets named G1 and G2
\cite{fur2}, do display naturalness and are not dominated by the last
terms retained; an evidence which confirms the usefulness of the EFT
concepts and validates the truncation of the effective Lagrangian at
the first lower orders.
The ideas of EFT have been fruitful \cite{fur3}, moreover, to
elucidate the empirical success of previous RMF models, like the
original $\sigma-\omega$ model of Walecka \cite{ser} and its
non-linear extensions with cubic and quartic scalar self-interactions
\cite{bog}. However, these conventional RMF models truncate the
effective Lagrangian at some level without further physical rationale
or symmetry arguments. The introduction of new interaction terms in
the effective model pursues an improved representation of the
relativistic energy functional \cite{fur2,ser2}.

Previous works have shown that the EFT model is able to describe in a
unified manner the properties of nuclear matter, both at normal and at
high densities \cite{est2,est}, as well as the properties of finite
nuclei near and far from the valley of $\beta$ stability
\cite{est3,hue}, with similar and even better quality to standard RMF
force parameters. With this positive experience at hand, in the
present paper we want to investigate the predictability of the new
effective Lagrangian approach to the nuclear many-body problem in
extrapolations to superheavy nuclei. Concretely, we shall focus on
analyzing the model predictions for spherical shell closures. Our
calculations will be performed in spherical symmetry. Though
deformation is an important degree of freedom for SHE
\cite{smo,bur,pat,ren_prc2,bur04}, we are searching for spherical
shell stability around $^{298}_{184}$114 and $^{292}_{172}$120 where
deformation is expected to be small and where the shell structure has
often been analyzed in the spherical approximation
\cite{smo,rut,ben,kru,rei,ben2}. For exploration, we also compute
hyperheavy nuclei around $N \sim 258$ which spherical calculations
have found to correspond to a possible region of increased shell
stability \cite{ben2}. Deformation would certainly change the picture
in the details and add deformed shell closures, e.g., like those
predicted around $Z=108$ and $N=162$ \cite{mol,smo,bur}, but it should
not change drastically the predictions for the values of $N$ and $Z$
where the strongest shell effects show up already in the spherical
calculation. Of course, for a quantitative discussion, one needs to
account for deformation effects which will serve to extend the island
of shell stabilized superheavy nuclei and to decide on the specific
form of the ground-state shapes of these nuclei.

Our analysis uses the EFT parameter sets G1 and G2 \cite{fur2}. The
results are compared with those obtained with the NL3 parameter set
\cite{lal3}, taken as one of the best representatives of the usual RMF
model with only scalar self-interactions. The paper is organized as
follows. In the second section we briefly summarize the RMF model
derived from EFT and our modified BCS approach to pairing. The third
section is devoted to the study of several properties of superheavy
nuclei such as two-particle separation energies and shell gaps,
average pairing gaps, single-particle energy spectra, and shell
corrections. The summary and conclusions are laid in the fourth
section. 

\section{Formalism}
\subsection{The model}

The EFT model used here has been developed in Ref.\ \cite{fur2}.
Further insight into the model and the concepts underlying it can be
gained from Refs.\ \cite{fur,ser2,fur3,rus}. For our purposes, the
basic ingredient is the EFT energy density functional for finite
nuclei. It reads \cite{fur2,ser2}
%
\begin{eqnarray}
\label{EDF}
{\cal E}({\bf r}) & = &  \sum_\alpha \varphi_\alpha^\dagger
\Bigg\{ -i \mbox{\boldmath$\alpha$} \!\cdot\! \mbox{\boldmath$\nabla$}
+ \beta (M - \Phi) + W
+ \frac{1}{2}\tau_3 R
+ \frac{1+\tau_3}{2} A
\nonumber \\[3mm]
& &
- \frac{i}{2M} \beta \mbox{\boldmath$\alpha$}\!\cdot\!
\left( f_v \mbox{\boldmath$\nabla$} W
+ \frac{1}{2}f_\rho\tau_3 \mbox{\boldmath$\nabla$} R
+ \lambda \mbox{\boldmath$\nabla$} A \right)
+ \frac{1}{2M^2}\left (\beta_s + \beta_v \tau_3 \right ) \Delta
A \Bigg\} \varphi_\alpha
\nonumber \\[3mm]
& & \null
+ \left ( \frac{1}{2}
+ \frac{\kappa_3}{3!}\frac{\Phi}{M}
+ \frac{\kappa_4}{4!}\frac{\Phi^2}{M^2}\right )
\frac{m_{s}^2}{g_{s}^2} \Phi^2  -
\frac{\zeta_0}{4!} \frac{1}{ g_{v}^2 } W^4
\nonumber \\[3mm]
& & \null + \frac{1}{2g_{s}^2}\left( 1 +
\alpha_1\frac{\Phi}{M}\right) \left(
\mbox{\boldmath $\nabla$}\Phi\right)^2
- \frac{1}{2g_{v}^2}\left( 1 +\alpha_2\frac{\Phi}{M}\right)
\left( \mbox{\boldmath $\nabla$} W  \right)^2
\nonumber \\[3mm]
& &  \null - \frac{1}{2}\left(1 + \eta_1 \frac{\Phi}{M} +
\frac{\eta_2}{2} \frac{\Phi^2 }{M^2} \right)
\frac{{m_{v}}^2}{{g_{v}}^2} W^2
- \frac{1}{2g_\rho^2} \left( \mbox{\boldmath $\nabla$} R\right)^2
- \frac{1}{2} \left( 1 + \eta_\rho \frac{\Phi}{M} \right)
\frac{m_\rho^2}{g_\rho^2} R^2
\nonumber \\[3mm]
& & \null
- \frac{1}{2e^2}\left( \mbox{\boldmath $\nabla$} A\right)^2
+ \frac{1}{3g_\gamma g_{v}}A \Delta W
+ \frac{1}{g_\gamma g_\rho}A \Delta R.
\end{eqnarray}
The coupling constants have been written so that in the present form
they should be of order unity according to the naturalness assumption.
The index $\alpha$ runs over all occupied nucleon states
$\varphi_\alpha ({\bf r})$ of the positive energy spectrum. The meson
fields are $\Phi \equiv g_{s} \phi_0({\bf r})$, $ W \equiv g_{v}
V_0({\bf r})$ and $R \equiv g_{\rho}b_0({\bf r})$, and the photon
field is $A \equiv e A_0({\bf r})$. Variation of the energy density
(\ref{EDF}) with respect to $\varphi^\dagger_\alpha$ and the meson
fields gives the Dirac equation fulfilled by the nucleons and the
Klein-Gordon equations obeyed by the mesons (see Refs.
\cite{fur2,est2} for the detailed expressions). We solve the Dirac
equation in coordinate space by transforming it into a
Schr\"odinger-like equation. 

In this work we shall employ the EFT parameter sets G1 and G2 of Ref.\
\cite{fur2} that were fitted by a least-squares optimization procedure
to 29 observables (binding energies, charge form factors and
spin--orbit splittings near the Fermi surface) of the nuclei $^{16}$O,
$^{40}$Ca, $^{48}$Ca, $^{88}$Sr and $^{208}$Pb. A satisfactory feature
of the set G2 is that it presents a positive value of $\kappa_4$, as
opposed to G1 and to most of the successful RMF parametrizations such
as NL3. We note that the value of the effective mass at saturation
$M^*_\infty/M$ in the EFT sets ($\sim 0.65$) is somewhat larger than
the usual value in the RMF parameter sets ($\sim 0.60$), which is due
to the presence of the tensor coupling $f_{v}$ of the $\omega$ meson
to the nucleon \cite{est3,fur4}. Also, the bulk incompressibility of
G1 and G2 is $K=215$ MeV, while the NL3 set has $K=271$ MeV\@.

\subsection{Pairing}

In order to describe open-shell nuclei the pairing correlations have
to be explicitly taken into account. The most popular approach for
well-bound isotopes has been the BCS method. However, the BCS
approximation breaks down for exotic nuclei near the drip lines
because it does not treat the coupling to the continuum properly. This
difficulty is disposed of either by the non-relativistic
Hartree-Fock-Bogoliubov theory, with Skyrme \cite{dob} and Gogny
\cite{dec} forces, or by the relativistic Hartree-Bogoliubov (RHB)
theory \cite{men,men2,lal4,vre}.

Pairing correlations are another important ingredient in the study of
superheavy elements. Furthermore, some of the predicted regions of
shell stability in superheavy nuclei lie close to the drip point and
a suitable treatment is required. Many calculations of SHE have often
used a zero-range two-body pairing force $V_{pair}=V_{0,p/n}\delta
({\bf r}-{\bf r}^\prime)$, with adjustable strengths for protons and
neutrons (see Refs. \cite{ben,rei}). A study of SHE using the RHB
approach, with the NL-SH parameter set, was carried out in Ref.\
\cite{lal}. 

To deal with the pairing correlations we use here a simplified
prescription which we have previously found to be in acceptable
agreement with RHB calculations \cite{est3}. The procedure is similar
to the one employed for Skyrme forces in Ref.\ \cite{cha}. For each
kind of nucleon we assume a constant pairing matrix element $G_q$,
which simulates the zero range of the pairing force, and we include
quasibound levels in the BCS calculation as done in Ref.\ \cite{cha}.
These levels of positive single-particle energy, retained by their
centrifugal barrier (neutrons) or by their centrifugal-plus-Coulomb
barrier (protons), mock up the influence of the continuum in the
pairing calculation. The wave functions of the considered quasibound
levels are mainly localized in the classically allowed region and
decrease exponentially outside it. As a consequence, the unphysical
nucleon gas which surrounds the nucleus if continuum levels are
included in the normal BCS approach is eliminated \cite{est3}. We
restrict the space of states involved in the pairing correlation to
one harmonic oscillator shell above and below of the Fermi level, to
avoid the unrealistic pairing of highly excited states and to confine
the region of influence of the pairing potential to the vicinity of
the Fermi level.

As described in Ref.\ \cite{est3}, the solution of the pairing
equations allows us to find the average pairing gap $\Delta_q$ for
each kind of nucleon. We write the pairing matrix elements as
$G_q=C_q/A$. We have fixed the constants $C_q$ by looking for the best
agreement of our calculation with the known experimental binding
energies of Ni and Sn isotopes for neutrons, and of $N=28$ and $N=82$
isotones for protons \cite{est3}. The values obtained from these fits
are $C_n=21$ MeV and $C_p=22.5$ MeV for the G1 set, $C_n=19$ MeV and
$C_p=21$ MeV for the G2 set, and $C_n=20.5$ MeV and $C_p=23$ MeV for
the NL3 interaction.

In Ref.\ \cite{est3} we applied this improved BCS approach
with the G1 and G2 parametrizations to study one- and two-neutron
(proton) separation energies for several chains of isotopes (isotones)
from stability to the drip lines. We found a reasonable agreement with
the available experimental data, similar to the one obtained using the
NL3 set. The analysis showed that the parameters sets based on EFT are
able to describe nuclei far from the $\beta$-stability valley when a
pairing residual interaction is included. 

\section{Results and discussion}

Traditionally a large gap in the single-particle spectrum has been
interpreted as an indicator of a shell closure, at least for nuclei of
atomic number $Z<100$. However, for a large nucleus like a superheavy
element, it may not be sufficient to simply draw the single-particle
level scheme and to look for the gaps, due to the complicated
structure of the spectrum  and the presence of levels with a high
degree of degeneracy.
Moreover, in a self-consistent calculation, a strong coupling between
the neutron and proton shell structure takes place.
Therefore, when dealing with SHE it is imperative to look for other
quantities to reliably identify the shell closures and magic numbers,
apart from the analysis of the single-particle level structure.

Here we shall consider the following observables as indicators for
shell closures:
\\
a) 
A sudden jump in the two-neutron (two-proton) separation energies of 
even-even nuclei, defined as
\begin{equation}
\label{s2n}
S_{2q}= E(N_q-2)-E(N_q),
\end{equation}
where $N_q$ is the number of neutrons (protons) in the nucleus
for $q= n$ ($q= p$).
A sharp drop in $S_{2q}$ means that a very small amount of energy is
required to remove two more nucleons from the remnant of the parent
nucleus. Thus, the parent nucleus is more stable which is a character
of magicity. This observable is an efficient tool to quantify the
shell effect because of the absence of odd-even effects \cite{ben}. 
\\
b) 
The size of the gap in the neutron (proton) spectrum is determined by
half of the difference in Fermi energy when going from a closed shell
nucleus to a nucleus with two additional neutrons (protons). This
quantity is very well accounted for by the two-neutron (two-proton)
shell gap which is defined as the second difference of the binding
energy \cite{rut,ben}: 
\begin{equation}
\label{D2Q}
\delta_{2q}(N_q)=S_{2q}(N_q)-S_{2q}(N_q+2)=E(N_q+2)-2E(N_q)+E(N_q-2).
\end{equation}
This quantity measures the size of the step found in the two-nucleon
separation energy and, therefore, it is strongly peaked at magic shell
closures. 
\\
c) 
The neutron and proton average pairing gaps $\Delta_q$ of open-shell
nuclei can be related to the odd-even mass difference, from where the
empirical law $\Delta \sim 12/\sqrt{A}$ can be derived \cite{rin}.
However, for closed shell nuclei $\Delta_q$ should vanish. Thus, we
shall use the vanishing of the average pairing gap obtained from
our calculations as another signal for identifying closed shell nuclei. 

We next calculate the above observables for the isotopic chain of
$Z=120$ and for several isotonic chains, assuming spherical symmetry.
We employ the parameter sets G1 and G2 due to the EFT formalism and
compare the results with those obtained from the standard RMF
parametrization NL3, which is well established as a successful
interaction for nuclei at and away from the line of $\beta$ stability.

It is to be mentioned that the previous indicators correspond to
energy differences between neighbouring nuclei. However, they do not
have a direct connection with the shell corrections which stabilize a
given ($N,Z$) superheavy nucleus against fission \cite{rei}. The shell
corrections are related to the difference between the nuclear binding
energies and the predictions of a liquid-drop model. As a
complementary study, after our search for spherical shell closures, we
shall analyze the shell corrections for the discussed chains of SHE.

\subsection{Isotopic chain of $\mbox{\boldmath$Z=120$}$}

We first consider the chain of isotopes with atomic number $Z=120$,
which is found as a magic number in recent relativistic mean field
calculations of nuclei in the superheavy mass region
\cite{rut,ben,kru,rei}. Figure 1 collects the results obtained with
the EFT parameter set G2. The two-neutron separation energies $S_{2n}$
are displayed in the upper panel of this figure. The $S_{2n}$ graph
shows a smooth decrease with increasing neutron number $N$ throughout
the whole chain except for the sudden jumps after the neutron numbers
$N=172$, 184, and 258. These jumps indicate the possible occurrence of
a shell closure at these neutron numbers. Using Eq.\ (\ref{D2Q}) we
calculate the two-nucleon shell gap for neutrons for the same isotopic
chain, and present the result in the middle panel of Figure 1. Sharp
peaks in $\delta_{2n}$ are found at the same neutron numbers 172, 184,
and 258. It is seen that the peaks of $N=172$ and $N=258$ are more
marked than the peak of $N=184$. Actually, the height of the peak at
$N=184$ is around only one third of that of the peak at $N=172$. One
may note that the amplitude of the jumps in $\delta_{2n}$ for shell
closures is smaller in the SHE region than in the region of normal
mass nuclei. This is expected due to the increase of the
single-particle level density with increasing mass number. For
example, for $^{208}$Pb and for the doubly-magic isotopes of tin and
calcium we find values of $\delta_{2n}$ between some 5 and 10 MeV\@.

The average pairing gap $\Delta_q$ is representative of the strength
of the pairing correlations. The curve for the neutron pairing gaps,
displayed in the bottom panel of Figure 1, shows a structure of arches
that vanish only at $N=172$, 184, and 258. Since the proton pairing
gap $\Delta_p$ is zero throughout the whole $Z=120$ chain, we have not
plotted it. Although we use a simplified prescription for the
calculation of the pairing gap \cite{est3}, our value for $\Delta_q$
can be considered as an average of the different state-dependent
single-particle gaps which would be obtained if one had used a
zero-range pairing force, as done e.g.\ in Refs.\ \cite{ben,rei}.

Therefore, all of the three analyzed observables are pointing to the
same neutron numbers as the best candidates for shell closures for
$Z=120$ with the G2 parametrization. The magic character of the proton
number $Z=120$ in combination with $N=172$, 184, and 258 is tested in
the calculations for isotonic chains that we present in the next
subsection. In analyzing the shell effects from a spherical
calculation it is to be kept in mind that only for doubly-magic nuclei
one can guarantee a spherical shape, when protons as well as neutrons
experience a spherical shell closure. For open-shell nuclei the
spherical solution does not always correspond to the ground state.
Inclusion of deformation might add extra stability for some $Z=120$
nuclei other than $^{292}120$, $^{304}120$ and $^{378}120$, and
perhaps additional peaks would develop with respect to the neighboring
background in the curve for $\delta_{2n}$, apart from the sudden jumps
we have detected by means of the spherical calculation. Only a
deformed calculation could definitively decide in such cases the
appropriate ground-state shape. Nevertheless, the spherical solution
gives a firsthand and overall view of the sequence of spherical shell
closures, which we have obtained from indicators which imply
differences of energies but not their absolute values.

In order to analyze the force dependence of the location of the shell
closures for the superheavy nuclei, we calculate the quantities
$S_{2n}$, $\delta_{2n}$ and $\Delta_n$ with the EFT set G1 and with
the NL3 parameter set for the same isotopic chain $Z=120$ and display
the results in Figures 2 and 3, respectively. As in the case of the G2
set, the proton pairing gap $\Delta_p$ vanishes for the whole chain of
$Z=120$ isotopes and it is not drawn. The global nature of the curves
of Figures 2 and 3 is quite similar to that observed previously with
the G2 set. Abrupt jumps in $S_{2n}$ and $\delta_{2n}$, and the
vanishing of the average neutron pairing gap $\Delta_n$, indicate
shell closures at $N=172$ and 258 in both of the G1 and NL3 sets. The
height of the peaks of $\delta_{2n}$ at $N=172$ and 258 is very
similar between the G1 and G2 sets, while they attain the largest
values in the NL3 set. For the $N=184$ system, $S_{2n}$ and
$\delta_{2n}$ show only a moderate jump in both G1 and NL3, indicating
a weaker shell closure than for $N=172$ and 258. Moreover, the neutron
pairing gap $\Delta_n$ does not vanish at $N=184$ with the G1 and NL3
sets, indicating that the occupancy of the single-particle levels is
diffused across the Fermi level, contrarily to the case of G2.

One expects a relatively large energy gap to appear between the last
occupied and the first unoccupied single-particle levels for the
neutron numbers corresponding to the shell closures detected above.
Let us now look into the neutron single-particle spectra, displayed in
Figure 4, for the $^{292}120$, $^{304}120$ and $^{378}120$ nuclei. All
the three parameter sets G2, G1 and NL3 show a large gap above the
Fermi energy for $N=172$ and 258. But for the neutron number $N=184$,
there appear only moderate gaps across the Fermi level for the G2 and
G1 sets, and the gap is still smaller for NL3. This is in agreement
with our previous discussions of the other indicators.

Inspecting the $N=258$ level spectrum one can appreciate another
visible energy gap across the neutron numbers $N=228$ (1k$_{17/2}$
level) and $N= 198$ (1j$_{13/2}$ level) in all of the parameter sets.
In spite of this, no distinct indications for a shell closure were
found for $N=228$ or $N=198$ in the curves of $S_{2n}$, $\delta_{2n}$,
and $\Delta_n$ in Figures 1--3. If one compares the spectra for the
three systems $N=172$, 184, and 258, it can be noticed that the gap
between two particular levels is strongly modified along the isotopic
chain. Consequently, an analysis of the spectra alone would not
suffice and the use of the discussed energy indicators becomes
mandatory in order to make predictions for shell closures in
superheavy nuclei.

\subsection{Isotonic chains}

We now proceed to discuss the isotonic chains of the neutron numbers
which we have detected as candidates for spherical shell closures in
the preceding study of the isotopic chain of $Z=120$. We start with
the $N=172$ isotonic chain in Figure 5, which displays the two-proton
separation energy $S_{2p}$, the two-proton shell gap $\delta_{2p}$,
and the average pairing gaps $\Delta_p$ and $\Delta_n$ in the
superheavy region from $Z=100$ up to the proton drip line, for the EFT
model G2 and for the conventional RMF model NL3. For brevity we do not
present the results from the G1 set, since the previous section has
shown that the predictions of G2 differ from NL3 more than in the case
of G1.

From Figure 5 one realizes that all the indicators signal a very
robust shell closure at $Z=120$, and a much weaker shell closure at
$Z=114$. The proton gap $\delta_{2p}$ ($\sim 5$ MeV) of the nucleus
$^{292}120$ is nearly twice as large as the corresponding neutron gap
$\delta_{2n}$ ($\sim 3$ MeV, Figures 1 and 3). For the NL3 set, in
addition, a little jump in $S_{2p}$ and a small peaked structure in
$\delta_{2p}$ indicates the possibility of a weak shell closure taking
place at $Z=106$. It is nevertheless known that the region around
$Z=106$ is deformed \cite{ben2} and thus the spherical solution does
not correspond to the ground state. Moreover, from the bottom panel of
Figure 5, we see that the neutron pairing gap $\Delta_n$ vanishes from
$Z=110$ till the proton drip point, but it is non-zero for smaller
atomic numbers. The non-vanishing neutron pairing gap for the
($N=172,Z=106$) combination tells us that this cannot be a
doubly-closed shell nucleus. Therefore, the neutron number $N=172$
exhibits a strong shell closure for the proton number $Z=120$ but this
magicity is washed out for $Z\le 110$. In conclusion, in the region of
superheavy nuclei the magicity of a particular neutron number depends
on the number of protons present in the nucleus.

We next analyze the shell closures for $N=184$ and $N=258$ in
combination with different proton numbers. In Figure 6 we display the
results for the $N=184$ isotonic chain. Curves are somewhat similar to
those for $N=172$. In the G2 parametrization one identifies $Z=114$
and $Z=120$ as possible shell closures though the size of the jump in
$\delta_{2p}$, which is similar for both proton numbers, is small if
we compare it with the jump observed for ($N=172,Z=120$) in Figure 5.
In the case of the NL3 parameter set there is some evidence for a weak
shell closure at $Z=114$ only, because the neutron pairing gap for
$Z=120$ does not vanish which prevents the combination ($N=184,Z=120$)
from representing a doubly-magic nucleus. The results for $N=258$
neutrons are shown in Figure 7. We again find a strong signature for a
shell closure at $Z=120$ in both the G2 and NL3 parameter sets,
whereas the indications for a shell closure at $Z=114$ are much
weaker. One also observes prominent jumps in $S_{2p}$ and
$\delta_{2p}$ at $Z=132$ (NL3) and $Z=138$ (NL3 and G2). But in this
region, close to the proton drip line, the neutron pairing gap does
not vanish. This implies that only the combination of $N=258$ with
proton numbers $Z=114$ and $Z=120$ may exhibit a double shell closure
character. 

Figure 8 depicts the proton single-particle spectra obtained with the
G2 and NL3 parametrizations for the illustrative examples of the
$^{286}114$, $^{292}120$, $^{298}114$, and $^{304}120$ nuclei. Looking
at the proton spectra for the systems with $N=172$, a very large gap
can be observed for 120 protons (between the 2f$_{5/2}$ and 3p$_{3/2}$
levels for G2, and between the 2f$_{5/2}$ and 1i$_{11/2}$ levels for
NL3). Instead, practically no gap exists for 114 protons (between the
2f$_{7/2}$ and 2f$_{5/2}$ levels), specially for the NL3 set. This is
consistent with the very weak signals of magicity of $Z=114$ in the
case of the $N=172$ isotonic chain shown by the $S_{2p}$ and
$\delta_{2p}$ indicators in Figure 5.

With the addition of only 12 neutrons, the proton spectra for the
systems with $N=184$ exhibit a different pattern than for $N=172$ near
the Fermi energy (cf.\ Figure 8). The gaps occurring between the
levels corresponding to 114 protons and to 120 protons are now
comparable in magnitude. This fact is in agreement with the relatively
magic character of the $^{298}114$ and $^{304}120$ nuclei predicted by
the indicators plotted in Figure 6. In any case, even for $N=184$, the
magicity of $Z=114$ is always smaller than the one shown by $Z=120$,
as one can see from the comparison of $S_{2p}$ and $\delta_{2p}$ in
Figures 5 and 6. This discussion shows again the strong dependence of
the proton (neutron) shell closures of SHE on the neutron (proton)
numbers and thus the importance of using the energy indicators.

\subsection{Shell corrections}

The stability of superheavy elements with an atomic number larger than
$Z \sim 100$ is possible thanks to the shell effects. In the liquid
droplet model picture these superheavy nuclei are unstable against
spontaneous fission because the large Coulomb repulsion can no longer
be compensated by the nuclear surface tension. However, SHE may still
exist because the quantal shell corrections generate local minima in
the nuclear potential energy surface which provide additional
stabilization. 

In our context the shell correction energy is also useful as a
different test for checking the robustness of the shell closures. For
experimentally known shell closures, i.e., up to $Z=82$ and $N=126$,
the shell corrections are strongly peaked around the magic numbers
(see, e.g., Ref.\ \cite{Kleban02}), providing enhanced binding for
magic nuclei. However, in the superheavy mass region, instead of
displaying sharp jumps, the shell corrections depict a landscape of
rather broad areas of shell stabilized nuclei \cite{kru,ben2}. Still,
in these areas the closed shell nuclei show a larger stabilization
(i.e., more negative shell corrections) than their neighbors. In the
present subsection we want to study the shell corrections around our
selected nuclei with $Z=114$ and $Z=120$, and $N=172$, $184$, and
$258$. 

The calculation of the shell correction energy is based on the
Strutinsky energy theorem \cite{str68} which states that the total
quantal energy can be divided in two parts:
\begin{equation}
\label{STU1}
E=\tilde{E}+E_{\rm shell}.
\end{equation}
The largest piece $\tilde{E}$ is the average part of the energy which
depends in a smooth way on the number of nucleons (namely, the part
well represented by the liquid droplet model). The smaller piece, the
shell correction $E_{\rm shell}$, has instead an oscillating
behaviour. The oscillations are due to the grouping of levels into
shells and display maxima at the shell closures.
According to the idea of Strutinsky, the average part of the
ground-state energy of a shell model potential can be obtained by
replacing the Hartree-Fock occupation numbers $n_\alpha$ (1 or 0 for
occupied or empty states) with occupation numbers $\tilde{n}_\alpha$
smoothed by an averaging function \cite{rin}. The shell correction
$E_{\rm shell}$ is computed as the difference of the exact energy
to that average part.

The Strutinsky smoothing procedure requires the use of several major
shells. This faces the problem of the treatment of the continuum when
realistic finite depth potentials are employed
\cite{kru,ben2,Bolster72,Nazar94}. Our strategy here, working in
coordinate space, and consistently with our approach to the treatment
of pairing, is to perform the Strutinsky smoothing including the
quasibound levels which are retained by their centrifugal barrier
(centrifugal-plus-Coulomb barrier for protons). We have taken 7 major
shells above the Fermi energy (i.e., states up to around 50 MeV above
the Fermi level) and have considered curvature corrections up to
$2M=10$ \cite{rin}. We have found that the plateau condition of the
averaged energy \cite{rin} is fulfilled for a smoothing parameter
$\gamma \sim 1.3-1.6$ MeV for both protons and neutrons. As we have
discussed, the quasibound levels included in our calculation do not
depend on the size of the box where the calculation is performed.
These levels, usually with high angular momentum, lie close in energy
to the RHB canonical levels \cite{est3}. Of course, one limitation of
our approach is that some resonant levels with low angular momentum
can be missed, more easily for neutrons, and then their contribution
is shared among the higher angular momentum levels which we include in
the calculation.

The total (neutron-plus-proton) shell corrections stemming from our
calculations for the isotopic chains with $Z=114$ and $Z=120$ are
displayed in Figure 9. The equivalent graph for the isotonic chains
with $N=172$, 184, and 258 is presented in Figure 10. Again, we point
out that our calculation is performed in spherical symmetry and thus
the calculated shell corrections represent in general an upper bound
to the actual ones. Stronger shell stabilization could still be
provided by deformation. The magnitude of the shell correction energy
$E_{\rm shell}$ is dictated by the level density around the Fermi
level. A high level density in the vicinity of the Fermi energy yields
a positive shell correction reducing the binding energy, whereas a low
level density gives a negative shell correction which increases the
binding energy. The shell corrections obtained with the G2 and NL3
sets are rather similar for the investigated isotopic and isotonic
chains. This is so because the single-particle levels around the Fermi
surface essentially show the same ordering with both parameter sets
and there are only small differences in the spin-orbit splittings, as
it can be realized from Figures 4 and 8. The results for the set G1
are also similar to those of G2 and NL3 and thus we do not display
them in Figures 9 and 10.

In Figure 9 the isotopic chain of $Z=120$ shows a large negative
shell correction at $N=172$, due to the presence of low angular
momentun levels near the Fermi energy for both neutrons (4s$_{1/2}$,
3d$_{3/2}$ and 3d$_{5/2}$ levels) and protons (3p$_{1/2}$ and
3p$_{3/2}$ levels). These levels imply a comparatively lower level
density and thus a more negative shell correction. The isotopic chain
also shows another local minimum around $N=182-184$, but in this case
the shell correction energy is less negative than for $N=172$. The
pattern exhibited by the total shell correction for the $Z=120$
isotopes looks very similar to that of the neutron shell correction
displayed in Figure 5 of Ref.\ \cite{kru} for the NL3 parameter set,
which was computed by means of the Green's function procedure. Looking
at the curves for the $Z=114$ isotopic chain represented in Figure 9
one realizes that the shell corrections are globally weaker than that
for $Z=120$ chain, which means less stability. They also present
minima at $N=172$ and at $N=184$, although in this case the situation
is reversed and the largest corrections correspond to $N=184$ instead
of $N=172$.

In the upper panel of Figure 10 the shell corrections for the $N=172$
isotonic chain clearly show only one minimum at $Z=120$. The $N=184$
chain (middle panel) displays one more local minimum at $Z=114$,
though the magnitude of the shell correction obtained for $Z=114$ is
smaller than for $Z=120$. 
 Our total shell corrections for $N=172$ and $N=184$ show similar
patterns to the proton shell corrections of NL3 which are depicted in
Figure 6 of Ref.\ \cite{kru} for these same isotonic chains.
The curves of the shell correction for the $N=258$ hyperheavy nuclei
(lower panel of Figure 10) are overall very much flat in comparison to
those for the isotonic chains of $N=172$ and $N=184$. The absolute
minimum again appears at $Z=120$. There is also a very small kink at
$Z=114$. The comparison of the curves in the 3 panels reveals that at
$Z=120$ all the curves show the most prominent minima. At $Z=114$, the
shell corrections for $N=172$ have no dip at all, but they display a
small kink for $N=184$ and $N=258$.

From the analysis of the shell correction energy we see that the
location of the minima in the shell stabilized regions of SHE is in
good agreement with the conclusions about the shell closures that we
inferred from the study of the previous indicators. Due to the fact
that the minima in the shell corrections for superheavy nuclei are
often not very pronounced, but rather shallow, we note the usefulness
of analyzing the shell corrections as a complementary means to assess
the predictions made on the basis of the energy indicators.

\section{Summary and conclusions}

We have investigated the predictions of the G1 and G2 parametrizations
of Ref.\ \cite{fur2} obtained from the modern effective field theory
approach to relativistic nuclear phenomenology for the occurrence of
spherical double shell closures and the shell stabilizing effect in
superheavy nuclei. Within an isotopic or isotonic chain of SHE the
possible shell closures are identified by a simultaneous occurrence at
a given $Z$ or $N$ of a large jump in the corresponding two-nucleon
separation energy $S_{2q}$, a pronounced peak in the two-nucleon shell
gap $\delta_{2q}$, and the vanishing of the average pairing gaps
$\Delta_n$ and $\Delta_p$. To treat the pairing correlations we have
employed an improved BCS model that was used successfully in Ref.\
\cite{est3} in calculations of isotopic and isotonic chains with magic
proton or neutron numbers.

First we have studied the isotopic chain of $Z=120$, which is found to
be a magic number in previous RMF calculations. Neutron shell closures
arise at $N=172$ and $N=258$ in all the considered parameter sets (G1,
G2 and NL3). In the particular case of the G2 set $N=184$ appears as
another possible neutron shell closure, though it is not as robust as
for $N=172$ or $N=258$. The magic character of $Z=120$ is supported by
the fact that the average proton pairing gap vanishes along the whole
isotopic chain. Next we have investigated the isotonic chains with
$N=172$, 184 and 258 for $Z > 100$. From this analysis the candidates
to proton shell closures have been found to be $Z=114$ (weakly) and
$Z=120$ (strongly). Other possible candidates different from $Z=114$
and $Z=120$ present a non-vanishing neutron pairing gap.
We conclude that the parameter sets G1 and G2 derived from the
effective field theory approach clearly point out towards the
doubly-magic character of the ($N=172,Z=120$) and ($N=258,Z=120$)
combinations, which is in agreement with the predictions of the NL3
set.

A minimum condition for a superheavy nucleus to be stable against
fission is that the shell effects must be able to provide enough
binding to compensate for the huge Coulomb repulsion among protons.
Compared to normal nuclei where the large negative shell corrections
are peaked at the magic numbers, the SHE display broad areas of shell
stabilization around the possible shell closures. We have computed the
shell corrections for the analyzed superheavy nuclei by means of an
Strutinsky smoothing. The continuum has been parametrized by taking
quasibound levels which in coordinate space are retained by their
centrifugal barrier.

We have found a region of shell stabilization for isotopes of
$Z=114$ and $Z=120$ in the range of neutron numbers $N \sim 170-186$.
The shell corrections are larger for $Z=120$ than for $Z=114$, and in
both cases show peaks at $N=172$ and $N=184$ which indicates the
larger stability of these nuclei relative to their neighbors. For the
isotonic chains of $N=172$ and $N=184$ the more negative shell
corrections appear at $Z=120$, although a smaller peak shows up at
$Z=114$ pointing out the relatively stable character of this nucleus,
at least compared with the immediate neighbors. The curve of the
shell corrections for the isotones of $N=258$ is mostly flat, but
again a depression can be recognized around $Z=120$. 

To summarize, in previous works \cite{est,est3} we showed that the
parameter sets derived from the effective field theory approach to the
low-energy nuclear many-body problem \cite{fur2} work nicely for both
$\beta$-stable and $\beta$-unstable nuclei. This is in addition to
their ability to yield a realistic equation of state at densities
above saturation which compares very favorably with microscopic
Dirac-Brueckner-Hartree-Fock calculations \cite{est}. In the present
study we have applied the EFT model to deal with the theoretical
description of some properties of superheavy nuclei. In particular, we
have seen that the G1 and G2 parameter sets reproduce the strong
double shell closure at $N=172$ and $Z=120$ predicted by the standard
RMF parametrizations, as well as a double shell closure at $N=258$ and
$Z=120$. Interestingly enough, the new parameter set G2 shows some
evidence for a double shell closure of the $N=184$, $Z=114$ nucleus
traditionally predicted by the macroscopic-microscopic models, as well
as by the Skyrme interaction SkI4.

The results presented here are a first prospect of the performance of
the EFT model in the description of SHE\@. Our calculations have been
restricted to spherical shapes. As we have discussed, this is not an
impeding drawback for the effects investigated in this work. However,
for comparisons with the measured data on the heavier actinides and
the experimentally synthesized SHE around $Z=110$, one definitively
needs to perform deformed calculations. In the future it will be
worthwhile trying to include deformation degrees of freedom into the
EFT model to extend the study to deformed nuclei of the SHE island.

\acknowledgments
Two of us (X. V. and M. C.) acknowledge financial support from the DGI
(Ministerio de Ciencia y Tecnolog{\'\i}a, Spain) and FEDER under grant
BFM2002-01868 and from DGR (Catalonia) under grant 2001SGR-00064. 
T. S. thanks the Spanish Education Ministry grant SB2000-0411 for
financial support and the Departament d'Estructura i Constituents de
la Mat\`eria of the University of Barcelona for kind hospitality.

%

\pagebreak

%
\section*{Figure captions}

\noindent
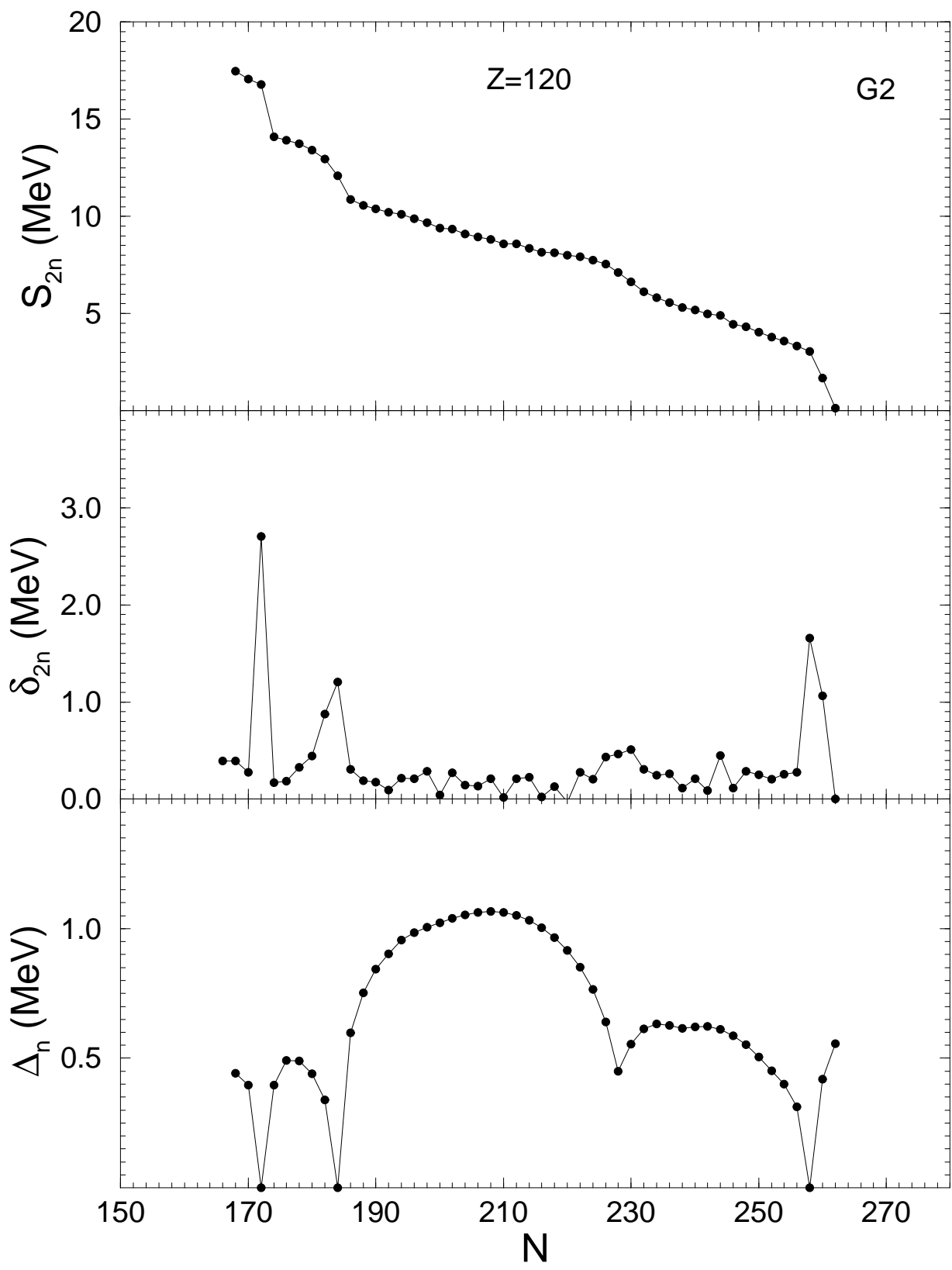
Figure 1.
The change with the neutron number $N$ of the two-neutron separation
energy $S_{2n}$, the two-neutron shell gap $\delta_{2n}$, and the
neutron average pairing gap $\Delta_n$ for $Z=120$ isotopes obtained
from spherical calculations with the relativistic parameter set G2.
The proton average pairing gap $\Delta_p$ vanishes in the whole
isotopic chain.
\\[2mm]
Figure 2.
Same as Figure 1 but for the relativistic parameter set G1.
\\[2mm]
Figure 3.
Same as Figure 1 but for the relativistic parameter set NL3.
\\[2mm]
Figure 4.
Single-particle spectrum of neutrons in the vicinity of the Fermi
level for the superheavy isotopes $^{292}120$, $^{304}120$, and
$^{378}120$ computed with the relativistic interactions G1, G2, and
NL3. 
\\[2mm]
Figure 5.
The change with the proton number $Z$ of the two-proton separation
energy $S_{2p}$, the two-proton shell gap $\delta_{2p}$, and the
proton $\Delta_p$ and neutron $\Delta_n$ average pairing gaps for
$N=172$ isotones obtained from spherical calculations with the
relativistic parameter sets G2 (left panels) and NL3 (right panels).
\\[2mm]
Figure 6.
Same as Figure 5 but for the isotones of $N=184$.
\\[2mm]
Figure 7.
Same as Figure 5 but for the isotones of $N=258$.
\\[2mm]
Figure 8.
Single-particle spectrum of protons in the vicinity of the Fermi level
for the superheavy isotones $^{286}114$ and $^{292}120$ (left panel),
and $^{298}114$ and $^{304}120$ (right panel) computed with the
relativistic interactions G2 and NL3.
\\[2mm]
Figure 9.
The change of the total shell correction energy (sum of the neutron
and proton contributions) with the neutron number $N$ for the isotopes
of $Z=114$ and $Z=120$ in spherical calculations with the G2 and NL3
models. 
\\[2mm]
Figure 10.
The change of the total shell correction energy (sum of the neutron
and proton contributions) with the proton number $Z$ for the isotonic
chains of $N=172$, $N=184$, and $N=258$ neutrons in spherical
calculations with the G2 and NL3 models. 
%
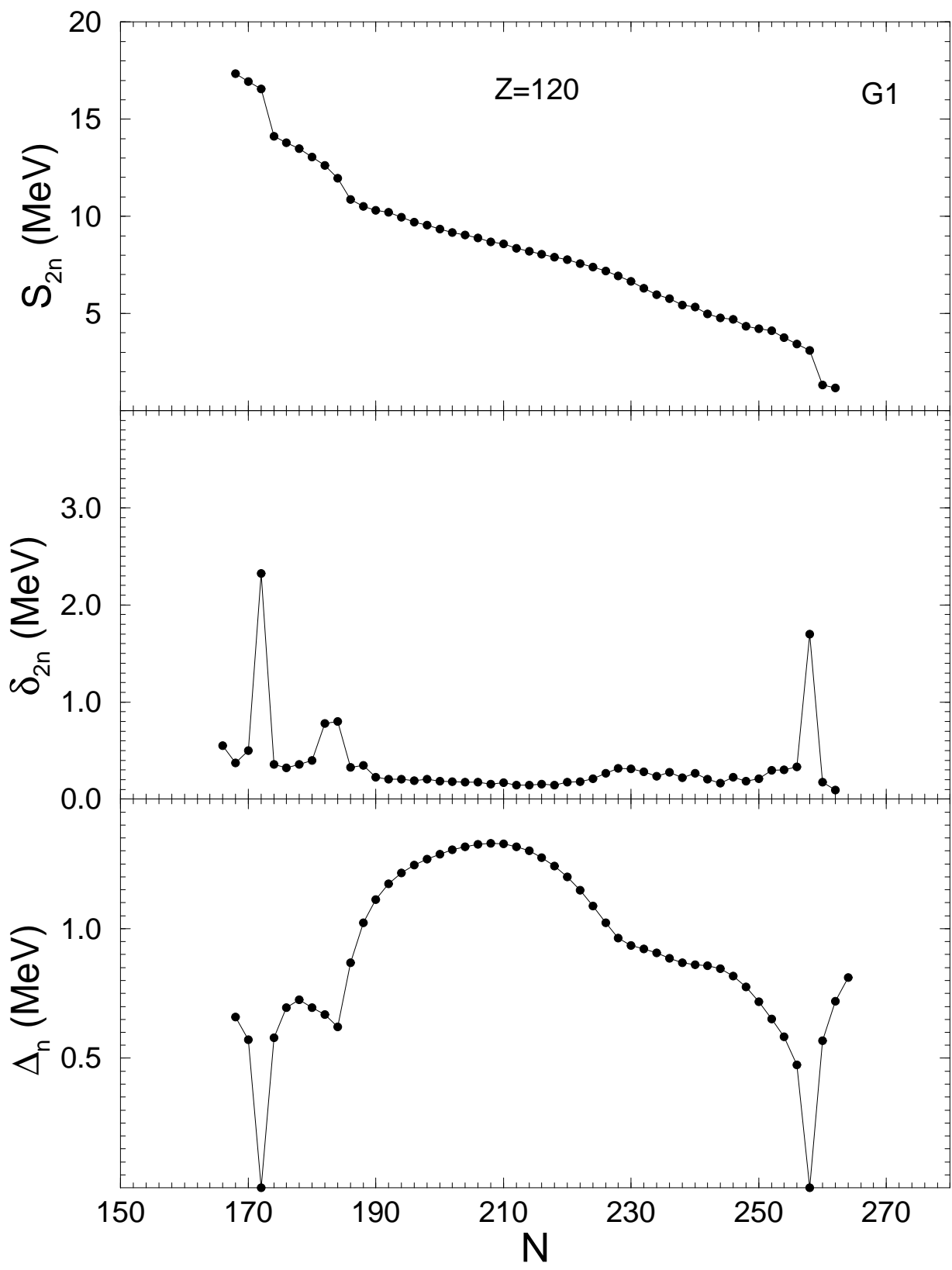
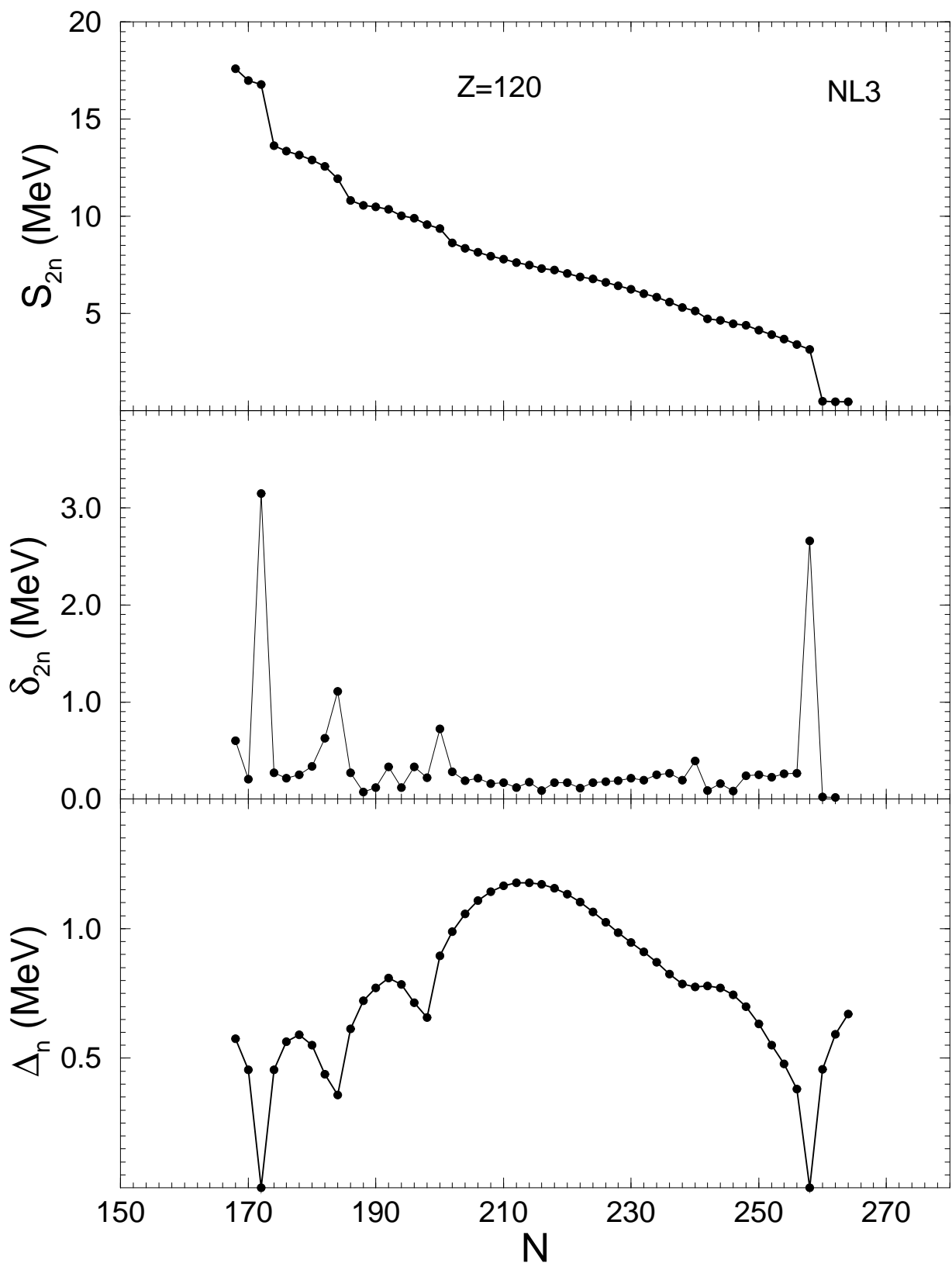
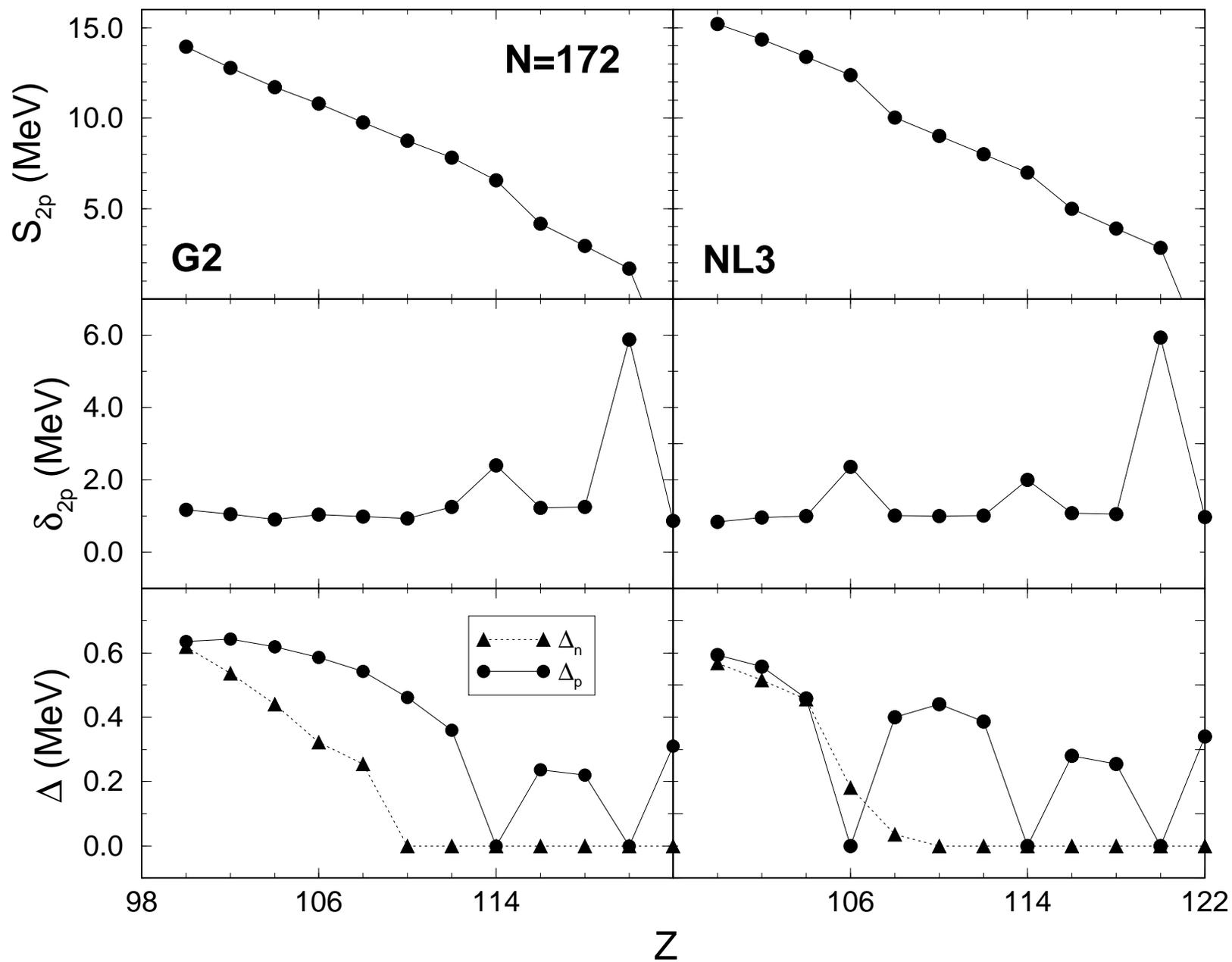
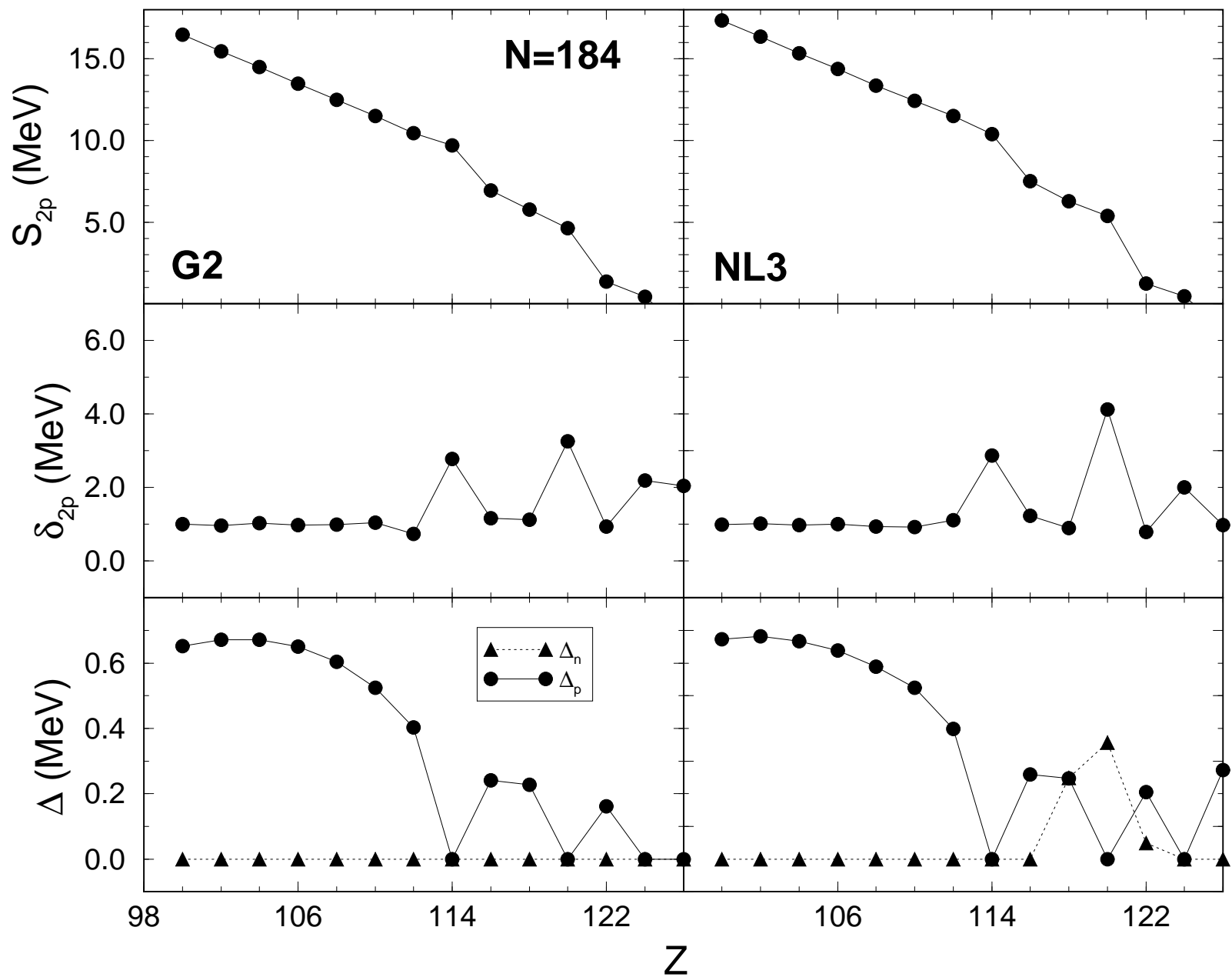
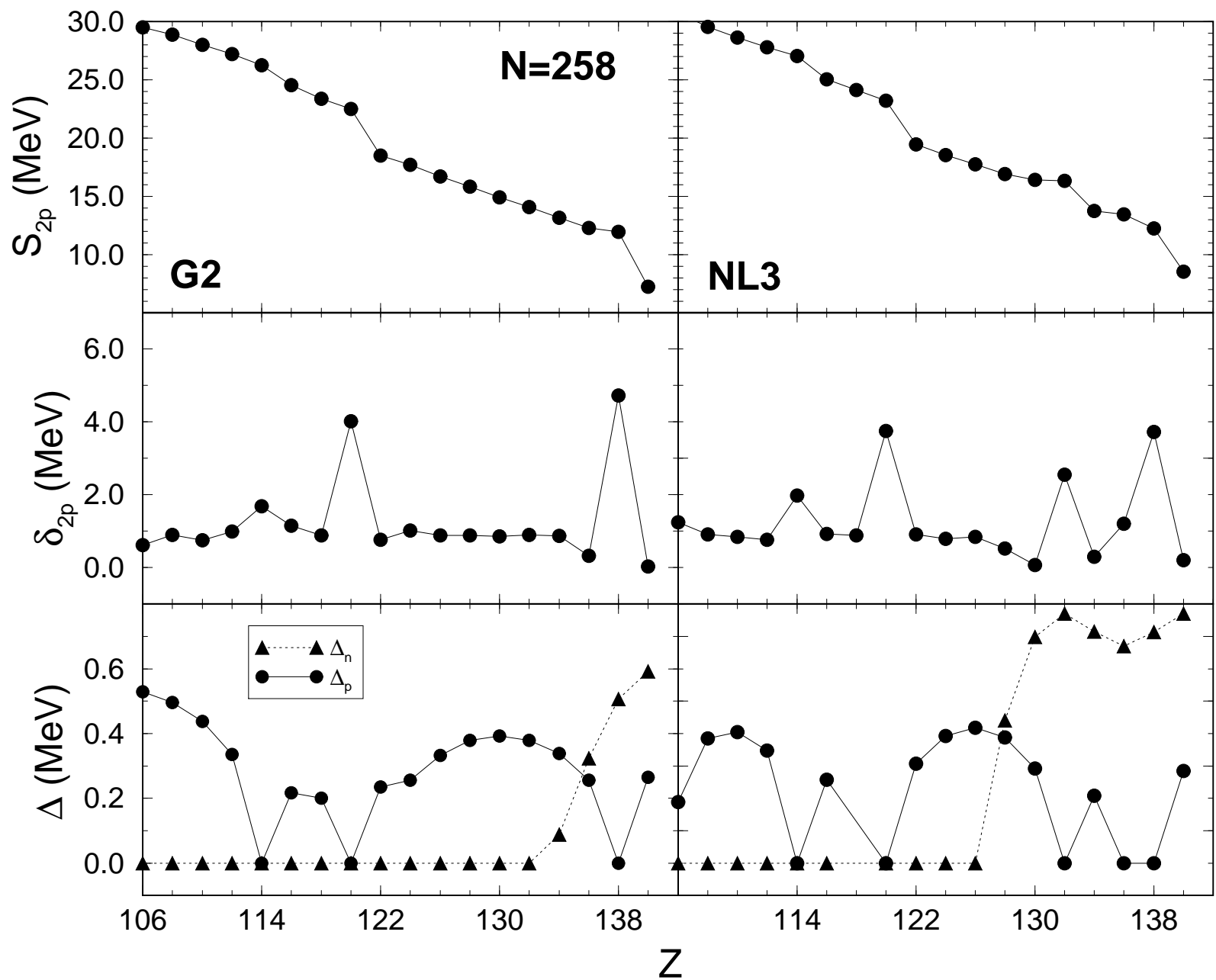
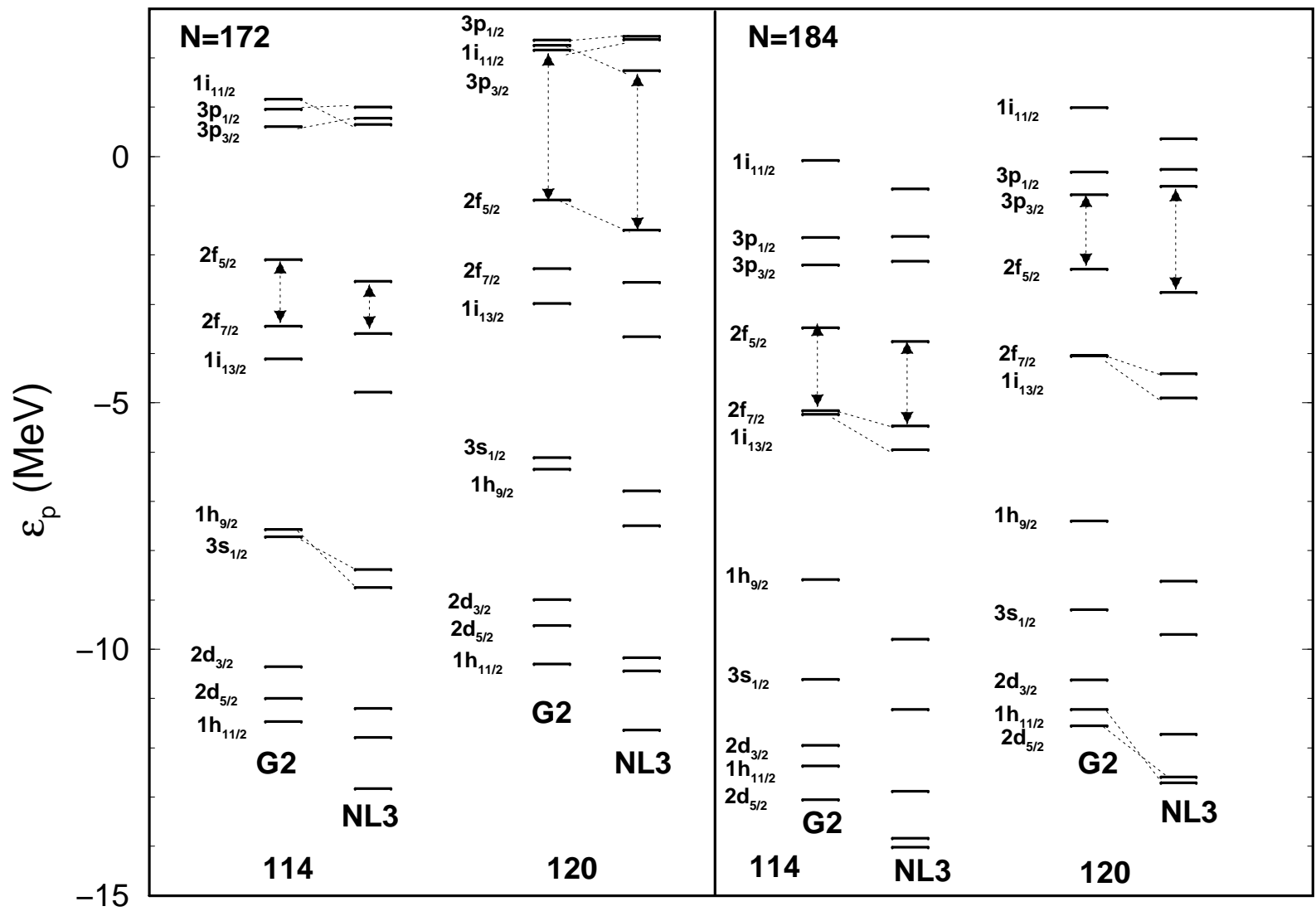
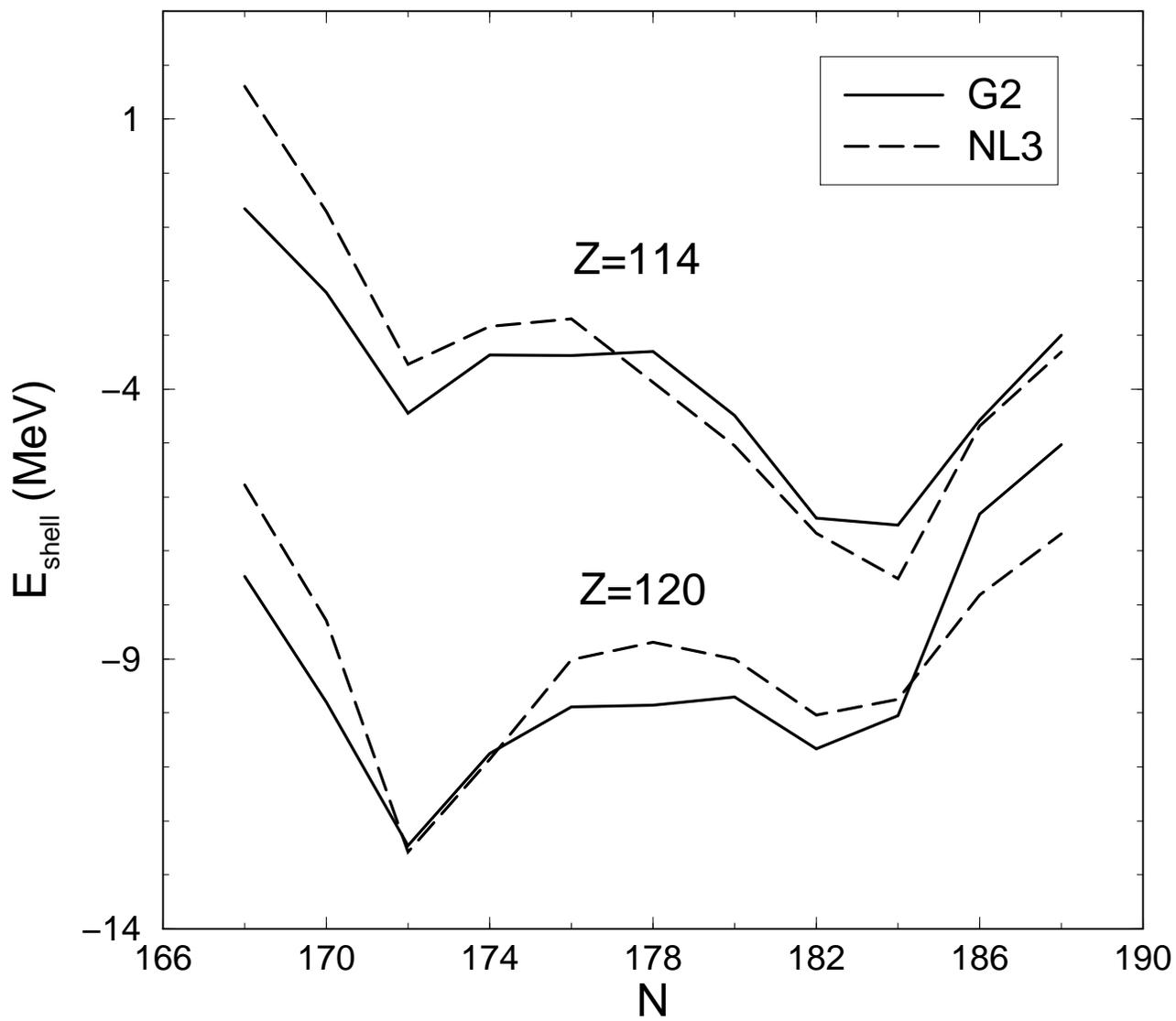
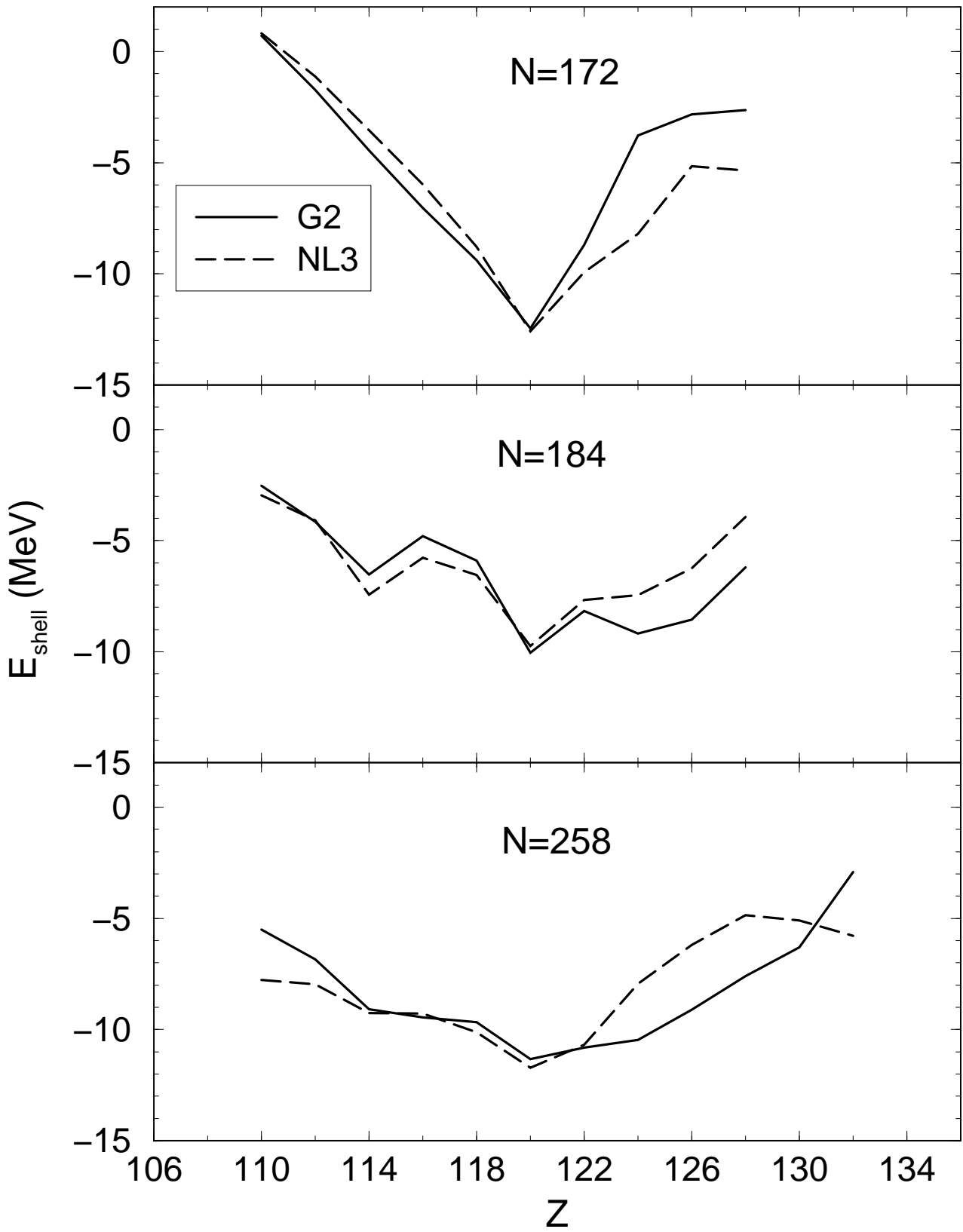
\end{document}